\documentclass[preprint, showkeys, showpacs, preprintnumbers, amsmath, amssymb, floatfix]{revtex4-1}
\usepackage{graphicx}% Include figure files
\usepackage{dcolumn}% Align table columns on decimal point
\usepackage{bm}% bold math

\begin{document}

\title{Anisotropic $H_{c2}$ of K$_{0.8}$Fe$_{1.76}$Se$_{2}$ determined up to 60 T}

\author{E. D. Mun, M.M. Altarawneh, C. H. Mielke, V. S. Zapf}
\affiliation{National High Magnetic Field Laboratory, Los Alamos National Laboratory, Los Alamos, New Mexico 87545, USA}
\author{R. Hu}
\author{S. L. Bud'ko}
\author{P. C. Canfield}
\affiliation{Ames Laboratory, US DOE and Department of Physics and Astronomy, Iowa State University, Ames, Iowa 50011, USA}

%\date{\today}

\begin{abstract}
The anisotropic upper critical field, $H_{c2}(T)$, curves for K$_{0.8}$Fe$_{1.76}$Se$_{2}$ are determined over a wide range of temperatures down to 1.5 K and magnetic fields up to 60 T.
Anisotropic initial slopes of $H_{c2}$ $\sim$ -1.4 T/K and -4.6 T/K for magnetic field applied along $\textbf{c}$-axis and $\textbf{ab}$-plane, respectively, were
observed. Whereas the $\textbf{c}$-axis $H_{c2}^{c}$ ($T$) increases quasi-linearly with decreasing temperature, the $\textbf{ab}$-plane $H_{c2}^{ab}$($T$) shows a flattening,
starting near 25 K above 30 T. This leads to a non-monotonic temperature dependence of the anisotropy parameter $\gamma_H\equiv$ $H_{c2}^{ab}/H_{c2}^{c}$. The anisotropy parameter is
$\sim 2$ near $T_c \sim 32$ K and rises to a maximum $\gamma_H\sim$ 3.6 around 27 K. For lower temperatures, $\gamma_H$ decreases with $T$ in a linear fashion, dropping to $\gamma_H \sim
2.5$ by $T \sim 18$ K. Despite the apparent differences between the K$_{0.8}$Fe$_{1.76}$Se$_{2}$ and (Ba$_{0.55}$K$_{0.45}$)Fe$_2$As$_2$ or Ba(Fe$_{0.926}$Co$_{0.074}$)$_2$As$_2$,  in terms of the magnetic state and proximity to an insulating state,
the $H_{c2}(T)$ curves are remarkably similar.
\end{abstract}

%%%%%%%%%%%%%%%%%%%%%%%%%%%%%%%%%%%%%%%%%%%%%%%%%%%%%%%%%%%%
%\keywords{Suggested keywords}%
%%%%%%%%%%%%%%%%%%%%%%%%%%%%%%%%%%%%%%%%%%%%%%%%%%%%%%%%%%%

\maketitle

Since the discovery of superconductivity in the FeAs-based family, intensive research efforts have focused on finding new Fe-based superconductors with a higher transition temperature,
$T_{c}$, and clarifying the pairing mechanism of superconductivity. \cite{Kamihara2008, Rotter2008, Wang2008, Hsu2008, Canfield2010, Paglione2010} As new families have been
discovered, the Fe-based superconductors have been categorized into several types including 11-type (P4/nmm, FeSe), 122-type (I4/mmm, AFe$_{2}$As$_{2}$, A = K, Sr, Ba), and 1111-type (P4/nmm, RFeAsO, R = rare
earth). Among these, FeSe is a simple binary system with $T_{c}\sim$ 8 K that can be increased up to 37 K by external pressure. \cite{Margadonna2009} Very recently, higher $T_{c}$
values ($> 30$ K) were successfully achieved by adding A (K, Rb, Cs, and Tl) \cite{Guo2010, Ying2010, Li2010, Fang2010} between the Fe$_{2}$Se$_{2}$ layers, changing the structure
from the 11-type (P4/nmm) to the 122-type (I4/mmm). The discovery of superconductivity in K$_{1-x}$Fe$_{2-y}$Se$_{2}$ \cite{Guo2010} materials, with the same ThCr$_{2}$Si$_{2}$
crystal structure as BaFe$_{2}$As$_{2}$ has given rise to a multitude of measurements, theories and questions. Even with the dramatically different transport properties (much closer to
semiconducting in temperature dependence and absolute value) and magnetic states (antiferromangetism existing up to above 500 K in samples that superconduct near 30 K)
\cite{Shermadini2011, Pomjakushin2011, Bao2011, Ryan2011} as compared to superconducting samples of doped BaFe$_{2}$As$_{2}$, the measured magnetic and transport anisotropy is modest \cite{Hu2011}
and $T_{c}$ values are comparable to K- and Co-doped members of the BaFe$_{2}$As$_{2}$ family. With these conspicuous differences, and clear similarities, it is of interest to establish
the anisotropic upper critical field, $H_{c2}$($T$), curves to see if they are similar or different from those found in the isostructural FeAs-based 122 materials. In this
communication, high magnetic field measurements of radio frequency (rf) contactless penetration depth into the mixed state are presented and the anisotropic $H_{c2}$ curves for
K$_{0.8}$Fe$_{1.76}$Se$_{2}$ are inferred. These results are compared with our earlier measurements of the anisotropic $H_{c2}$($T$) curves for K- and Co-doped BaFe$_{2}$As$_{2}$.
\cite{Altarawneh2008, Ni2008, Yuan2009, Kano2009}

The single crystals of K$_{0.8}$Fe$_{1.76}$Se$_{2}$ were grown from a high temperature melt and the actual composition was determined by the Wavelength Dispersive X-ray Spectroscopy
(WDS). Details of sample growth and characterization are given in Ref. \cite{Hu2011}. The temperature and magnetic field dependences of the electrical resistance were measured using
the four probe ac ($f$ = 16 Hz) technique in a Physical Property Measurement System (Quantum Design). To investigate the upper critical field anisotropy to higher fields ($H \leq 60$
T), the magnetic field dependence of radio frequency (rf) contactless penetration depth was measured for applied field both parallel ($\textbf{H}\parallel\textbf{c}$) and
perpendicular ($\textbf{H}\parallel\textbf{ab}$) to the tetragonal \textbf{c}-axis. The rf contactless penetration depth measurements were performed in a 60 T short pulse magnet with
a 10 ms rise and 40 ms decay time. The rf technique has proven to be a sensitive and accurate method for determining the $H_{c2}$ of superconductors. \cite{Mielke2001} This technique
is highly sensitive to small changes in the rf penetration depth ($\sim$ 1-5 nm) in the mixed state. As the magnetic field is applied, the probe detects the transition to
the normal state by tracking the shift in resonant frequency, which is proportional to the change in penetration depth as $\Delta \lambda \propto \Delta F/F_{0}$, where $F_{0}$ is 25
MHz in the current setup. Because of the eddy current heating caused by the pulsed field, small single crystals were chosen, where the sample was placed in a circular detection coil for
$\textbf{H}\parallel\textbf{ab}$ and was located on the top surface of one side of the counterwound coil pair for $\textbf{H}\parallel\textbf{c}$. \cite{Coffey2000, Altarawneh2009}
For the $\textbf{H}\parallel\textbf{c}$ configuration, the coupling between sample and coil is weaker than that for $\textbf{H}\parallel\textbf{ab}$, resulting in a smaller frequency shift that is still sufficient to resolve $H_{c2}(T)$. Details about this technique can be found in Refs. \cite{Altarawneh2008, Coffey2000, Altarawneh2009}.

\begin{figure}
\centering
\includegraphics[width=1.0\linewidth]{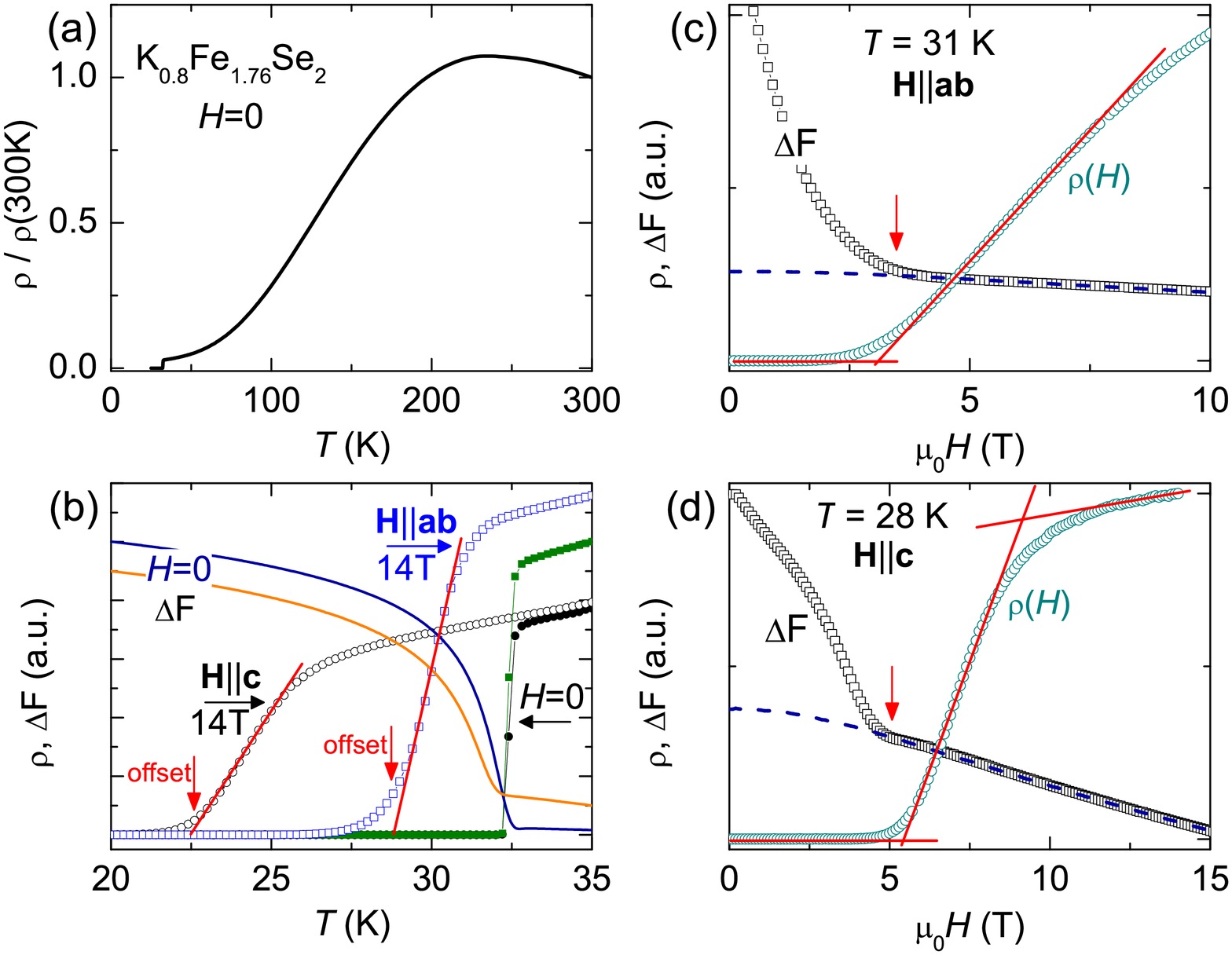}
\caption{(a) Temperature dependence of the normalized $\textbf{ab}$-plane resistivity $\rho(T)$ of the K$_{0.8}$Fe$_{1.76}$Se$_{2}$ single crystal at $H$ = 0, where $\rho$(300K) =
0.12 $\Omega$ cm. \cite{Hu2011}  (b) Low temperature region of the resistance for two samples at $H$ = 0 (closed symbols) and 14 T (open symbols) and the warming curves of rf shift
($\Delta$F) for two samples (solid lines). Vertical arrows indicate $T_{c}^{\texttt{offset}}$ and lines on the top of 14 T data are guide to the eye. (c) Comparison of the
$\textbf{ab}$-plane resistance $R(H)$ and $\Delta$F for $\textbf{H}\parallel\textbf{ab}$ at $T$ = 31 K. (d) Comparison of the $\textbf{ab}$-plane resistance $R(H)$ and $\Delta$F for
$\textbf{H}\parallel\textbf{c}$ at $T$ = 28 K. The dashed lines in (c) and (d) are the $\Delta$F taken at $T$ = 35 K as a normal state, background signal. The solid lines in (c) and (d) are guides to
the eye for offset and onset criteria of $H_{c}$ and vertical arrows indicate the deviation of $\Delta F$ from the background signal (see text).}
\label{Fig1}%
\end{figure}

Figure \ref{Fig1}(a) shows the temperature dependence of the normalized resistivity for the K$_{0.8}$Fe$_{1.76}$Se$_{2}$ sample. A sharp drop, corresponding to the superconducting
transition, was observed around 32 K. At high temperatures, the resistance increases with decreasing temperature and exhibits a broad maximum at around 220 K. \cite{Hu2011,
Mizuguchi2011} The offset and zero-resistance ($R < 3\times10^{-5} \Omega$) temperatures were estimated to be $T_{c}^{\texttt{offset}}$ $\simeq$ 32.2 K and $T_{c}^{\texttt{zero}}$
$\simeq$ 32 K, respectively, as shown in Fig. \ref{Fig1}(b). The solid lines in Fig. 1 (b) are warming curves of the rf shift ($\Delta$F) at $H$ = 0 for two different samples. As the
temperature decreases, the rf shift suddenly increases at $T_{c}$, where $T_{c}$ = 32 and 32.4 K for two samples were determined from d$\Delta$F/d$T$. A clear anisotropy in the
response of the superconductivity under applied fields was observed between $\textbf{H}\parallel\textbf{ab}$ and $\textbf{H}\parallel\textbf{c}$ as shown in Fig. 1(b) for $H$ = 14 T
curves.  To compare the superconducting transition between resistance and $\Delta$F measurement, resistance data measured in a superconducting magnet and $\Delta$F taken in pulsed
magnetic fields at $T$ = 31 K for $\textbf{H}\parallel\textbf{ab}$ and at $T$ = 28 K for $\textbf{H}\parallel\textbf{c}$ are plotted in Figs. 1(c) and (d), respectively. As shown in
the figures, the deviation from the background signal of $\Delta$F is close to the $H_{c}^{\texttt{offset}}$ criterion of the resistance curves.

\begin{figure}
\centering
\includegraphics[width=1.0\linewidth]{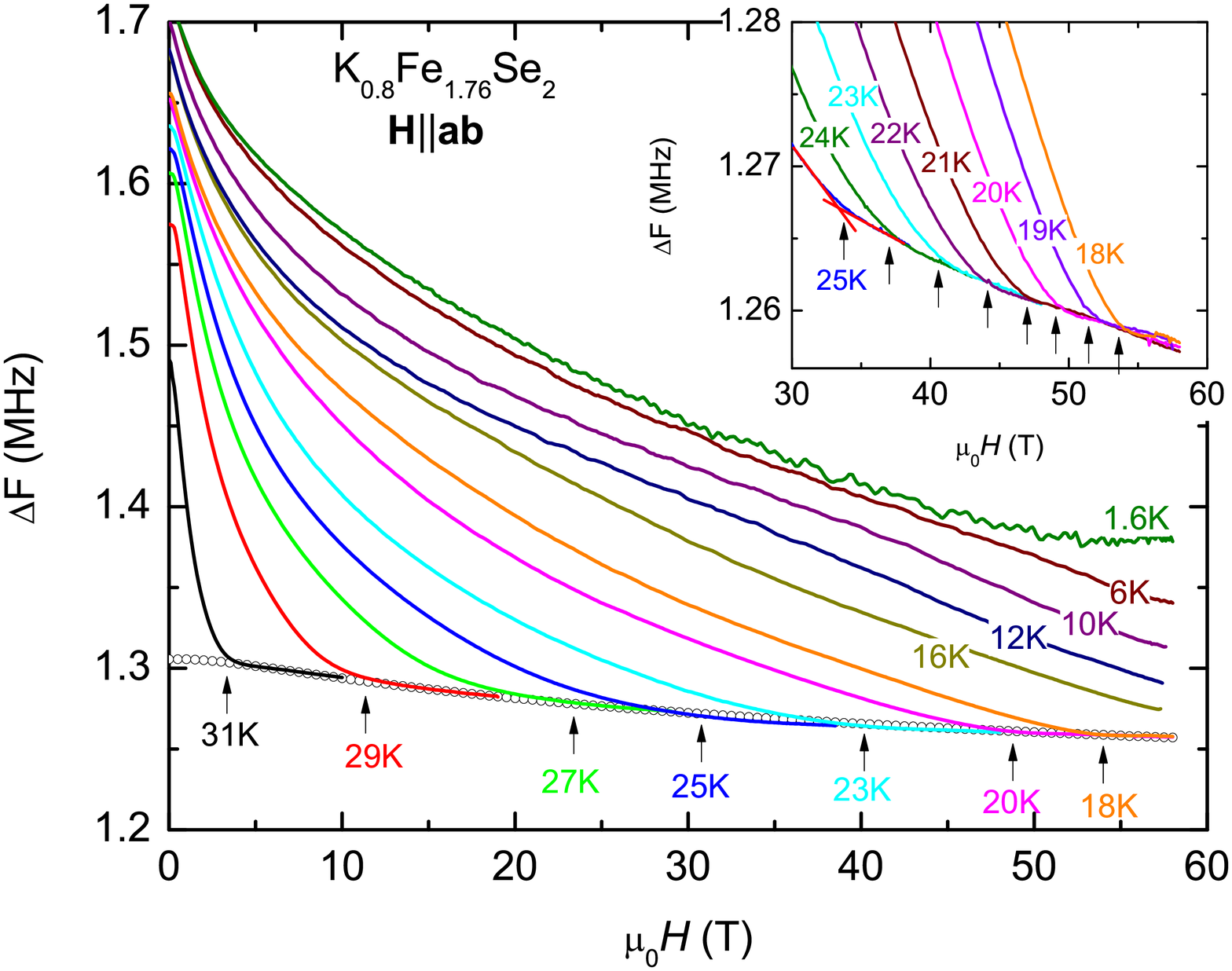}
\caption{Frequency shift ($\Delta$F) as a function of magnetic field for $\textbf{H}\parallel\textbf{ab}$ at selected temperatures. Open symbols are $\Delta$F taken at $T$ = 35 K as a normal state,
background signal. The arrows indicate $H_{c2}$ determined from the point deviating from background signal. Inset shows the low temperature data close to $H_{c}$. The straight lines on
the $T$ = 25 K curve are guides to the eye for determining the point at which the rf signal intercepts the slope of the normal state background.}
\label{Fig2}%
\end{figure}

\begin{figure}
\centering
\includegraphics[width=1.0\linewidth]{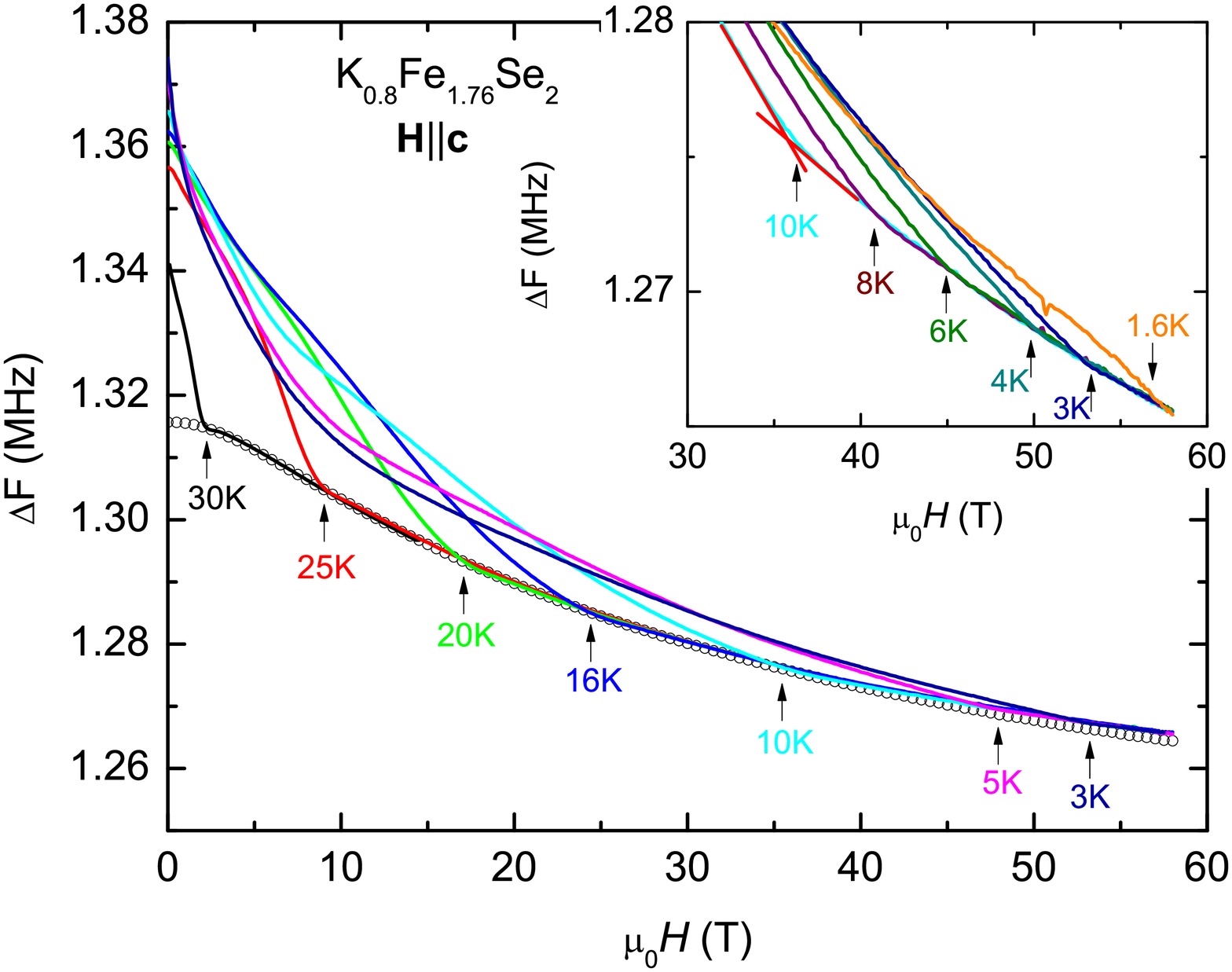}
\caption{Frequency shift ($\Delta$F) as a function of magnetic field for $\textbf{H}\parallel\textbf{c}$ at selected temperatures. Open symbols are $\Delta$F taken at $T$ = 35 K as a normal state,
background signal. The arrows indicate $H_{c2}$ determined from the point deviating from background signal. Inset shows the low temperature data close to $H_{c}$. The straight lines
on $T$ = 10 K curve are guides to the eye for determining the point at which the rf signal intercepts the slope of the normal state background.}
\label{Fig3}%
\end{figure}

Using the deviation from normal state criterion just discussed, the $\Delta$F vs $H$ plots shown in Figs. \ref{Fig2} and \ref{Fig3} can be used to infer the temperature dependence of
the upper critical field $H_{c2}(T)$ by taking the slope of the rf signal intercepting the slope of the normal state background or by simply taking the first point deviating from the
normal state background. Importantly, the difference between these two related criteria is small and does not affect $H_{c2}$($T$) curve. In the high temperature region, the point at
which the $\Delta$F signal deviates from the background is close to the $H_{c}^{\texttt{offset}}$ of the resistance data as shown in Fig. \ref{Fig1} (c) and (d). Therefore, $H_{c2}$
was determined at the point at which $\Delta$F deviates from the background signal. Arrows in Figs. \ref{Fig2} and \ref{Fig3} indicate the determined $H_{c2}$. The difference between the $H_{c2}$ values determined by the first deviation and slope change point criteria was used to determine the $H_{c2}$ error bar size.

\begin{figure}
\centering
\includegraphics[width=1.0\linewidth]{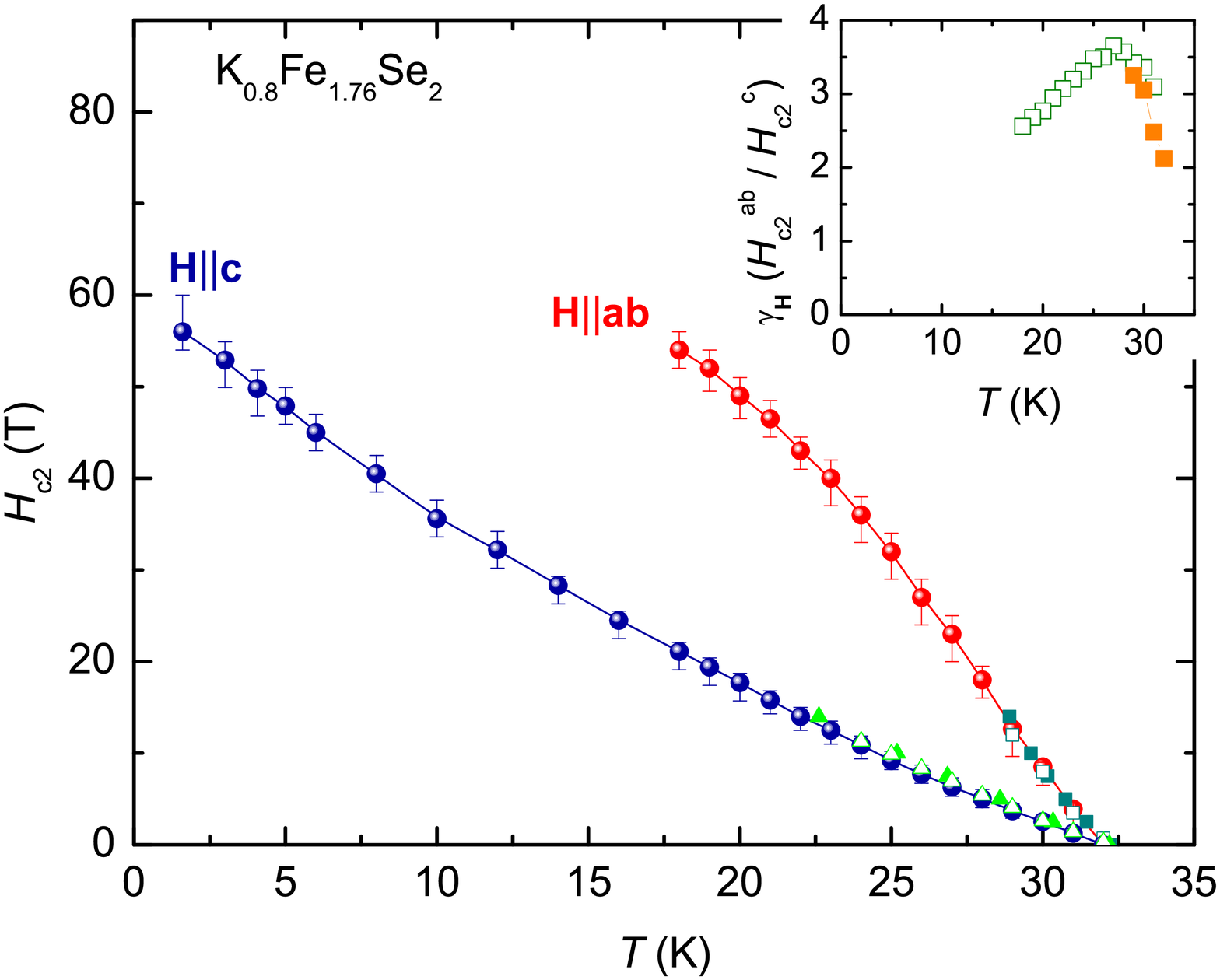}
\caption{Anisotropic $H_{c2}$($T$) for K$_{0.8}$Fe$_{1.76}$Se$_{2}$ single crystals. Solid circles are obtained from the pulsed field rf shift measurements and closed (open) square
and triangle symbols are taken from temperature (magnetic field)-dependent resistance measurements. Inset shows the temperature dependence of the anisotropy $\gamma_H$ =
$H_{c2}^{ab}/H_{c2}^{c}$ as determined from pulsed field rf shift (open squares) and resistance (solid squares) measurements.}
\label{Fig4}%
\end{figure}

The $H_{c2}$($T$) curves for both $\textbf{H}\parallel\textbf{ab}$ ($H_{c2}^{ab}$) and $\textbf{H}\parallel\textbf{c}$ ($H_{c2}^{c}$) in K$_{0.8}$Fe$_{1.76}$Se$_{2}$ are plotted in
Fig. \ref{Fig4}, as determined from the $H \leq 14$ T resistance and from the $H \leq 60$ T data taken from the down sweep of pulsed field magnetic field rf measurements. The
curvature of $H_{c2}$($T$) has been reported to vary depending on the criteria used to determine $H_{c2}$, for example in the case of highly two dimensional, high-$T_{c}$ cuprate
superconductors. \cite{Ando1999} In this study, the shape of $H_{c2}$ curves does not change qualitatively when $H_{c2}$ is defined by different criteria or even different
measurements. On the other hand, the shapes of the upper critical field curves for $\textbf{H}\parallel\textbf{ab}$ and $\textbf{H}\parallel\textbf{c}$ clearly do not manifest the
same temperature dependence. As is evidenced from Fig. \ref{Fig4}, a conventional linear field dependence of $H_{c2}$ is observed close to the $T_{c}$, with clearly different slopes
for the two field orientations. In the low field region the $H_{c2}$ curves are consistent with earlier studies. \cite{Mizuguchi2011, Wang2011} Towards higher fields, $H_{c2}^{c}$($T$) presents
an almost linear temperature dependence down to 1.5 K, whereas the curve of $H_{c2}^{ab}$($T$) has a tendency to saturate. The anisotropy parameter, $\gamma_H\equiv$
$H_{c2}^{ab}/H_{c2}^{c}$, is about $\sim$ 2 near $T_{c}$, but shows a maximum around 27 K with $\gamma_H\sim$ 3.6, and decreases considerably for lower temperatures. In all known examples so far, the temperature dependence of $\gamma_H$ was opposite to that of $\gamma_\lambda \equiv \lambda_c/\lambda_{ab}$. It would be interesting to examine $\gamma_\lambda(T)$ in this material, in particular to see if it goes through a minimum at $\sim 27$ K.

The zero temperature limit of $H_{c2}$ can be estimated by using the Werthamer-Helfand-Hohenberg (WHH) theory, \cite{Werthamer1966} which gives $H_{c2}$ =
0.69$T_{c}$(d$H_{c2}$/d$T$)$\mid_{T_{c}}$. The value of $H_{c2}$(0) for $\textbf{H}\parallel\textbf{ab}$ and $\textbf{H}\parallel\textbf{c}$ were estimated to be $\sim$ 102 T and $\sim$
31 T respectively, where $T_{c}$ = 32 K, d$H_{c2}^{ab}$/d$T$ $\sim$ -4.6 T/K and d$H_{c2}^{c}$/d$T$ $\sim$ -1.4 T/K were used. Clearly these values do not capture the salient physics
for this compound. On the other hand, in the simplest approximation, the Pauli limit ($H_{P}$) is given by 1.84$T_{c}$, \cite{Clogston1962, Chandrasekhar1962, Maki1966} giving $H_{P}$
$\sim$ 59 T. This low temperature value of $H_{c2}$ may indeed capture some of the basic physics associated with K$_{0.8}$Fe$_{1.76}$Se$_{2}$. To explain the observed $H_{c2}$ curves
in detail, a more complete theoretical treatment is needed, one that does not exclude the strong electron-phonon coupling and multiband nature of Fe-based compounds.  Anisotropic superconducting coherence length can be calculated using $H_{c2}^{ab} = \frac{\phi_0}{2 \pi \xi_{ab} \xi_c}$ and  $H_{c2}^{c} = \frac{\phi_0}{2 \pi \xi_{ab}^2}$. \cite{Poole2000} If $H_{c2}^c = 60$ T and $H_{c2}^{ab}$ is assumed to be between 60 and 100 T, then $\xi_{ab} \sim 2.3$ nm, and 1.4 nm $\lesssim \xi_c \lesssim 2.3$ nm.

On the basis of this study for K$_{0.8}$Fe$_{1.76}$Se$_{2}$, the behavior of $H_{c2}$($T$) is found to be very similar to that of several 122 systems as well as doped FeSe.
\cite{Altarawneh2008, Ni2008, Yuan2009, Kano2009} It should be noted that the $H_{c2}$ curves for two orientations in K-doped BaFe$_{2}$As$_{2}$ system seem to cross at low
temperature due to the flattening of $H_{c2}^{ab}$($T$) curve, \cite{Altarawneh2008, Yuan2009} additionally, the $H_{c2}$ curves for FeTe$_{0.6}$Se$_{0.4}$ shows a crossing between
$\textbf{H}\parallel\textbf{ab}$ and $\textbf{H}\parallel\textbf{c}$ curves below 4.5 K because of the subsequent flattening of the $H_{c2}^{ab}$ curve at low temperatures. \cite{Khim2010,
Fang2010R} However in the Co-doped system, the anisotropic $H_{c2}(T)$ curves do not show such crossing, \cite{Ni2008, Kano2009} a result similar to what was found in this
study. Thus, an intriguing feature of $H_{c2}$($T$) curves for Co- and K-doped BaFe$_{2}$As$_{2}$, FeTe$_{0.6}$Se$_{0.4}$ and K$_{0.8}$Fe$_{1.76}$Se$_{2}$ systems is that the
anisotropy near $T_c$ is as large as 3 but drops towards $\sim$ 1 as $T \rightarrow 0$ K. The $H_{c2}(T)$ anisotropy in K$_{0.8}$Fe$_{1.76}$Se$_{2}$ is particularly noteworthy given
that it exists deep within an antiferromagnetically ordered state. \cite{Pomjakushin2011, Bao2011,  Ryan2011} In the case of Co-doped Ba122, $\gamma_H(T) \sim 1$ when $T_c < T_N$ with
clear anisotropy only emerging when the antiferromagnetic state is suppressed. \cite{Ni2008} These results raise the question, to what extent is the antiferromagnetism in
K$_{0.8}$Fe$_{1.76}$Se$_{2}$ interacting with lower temperature superconductivity? Clearly more work will be needed to answer this key query.

{\it In summary}, the $H_{c2}$-$T$ phase diagram for the K$_{0.8}$Fe$_{1.76}$Se$_{2}$ has been constructed by means of measuring both the electrical resistance in a dc superconducting magnet ($H
\leq 14$ T) and the radio frequency contactless penetration depth in pulsed magnetic field up to 60 T. The upper critical field of K$_{0.8}$Fe$_{1.76}$Se$_{2}$ is determined as
$H_{c2}^{ab}$(18 K) $\simeq$ 54 T and $H_{c2}^{c}$(1.6 K) $\simeq$ 56 T. The anisotropy parameter $\gamma_H$ initially increases with decreasing temperature, passed through a maximum of
$\sim 3.6$ near 27 K, then decreases to $\sim 2.5$ at 18 K. The observed $\gamma_H$ values show a weakening anisotropic effect at low temperatures. Although the Fe-based superconductors
have a layered crystal structure, a weak anisotropy of $H_{c2}$ may be a common feature, suggesting that the inter-layer coupling and the three dimensional Fermi surface may play an
important role in the superconductivity of this family.

\begin{acknowledgments}
We thank V. G. Kogan for edifying and uplifting discussions. Work at the NHMFL-PFF is supported by the NSF, the DOE and the State of Florida. R.H. and P.C.C. are supported by AFOSR MURI grant \#FA9550-09-1-0603. S.L.B. was supported in part by the State of Iowa through the Iowa State University. S.L.B. and P.C.C. are also
supported by the U.S. Department of Energy, Office of Basic Energy Science, Division of Materials Sciences and Engineering. Ames Laboratory is operated for the U.S. Department of
Energy by Iowa State University under Contract No. DE-AC02-07CH11358. 
\end{acknowledgments}

\clearpage

\begin{thebibliography}{99}

\bibitem{Kamihara2008}%LaFeAsO
Y. Kamihara, T. Watanabe, M. Hirano, and H. Hosono, J. Am. Chem. Soc. {\bf 130}, 3296 (2008).

\bibitem{Rotter2008}%BaFe2As2
M. Rotter, M. Tegel, and D. Johrendt, Phys. Rev. Lett. {\bf 101}, 107006 (2008).

\bibitem{Wang2008}%LiFeAs
X. C. Wang, Q. Q. Liu, Y. X. Lv, W. B. Gao, L. X. Yang, R. C. Yu, F. Y. Li, and C. Q. Jin, Solid State Commun. {\bf 148}, 538 (2008).

\bibitem{Hsu2008}%FeSe
F. C. Hsu, J. Y. Luo, K. W. Yeh, T. K. Chen, T. W. Huang, P. M. Wu, Y. C. Lee, Y. L. Huang, Y. Y. Chu, D. C. Yan, and M. K. Wu, Proc. Natl. Acad. Sci. U.S.A. {\bf 105}, 14262 (2008).

\bibitem{Canfield2010}%review
Paul C. Canfield and Sergey L. Bud'ko, Annual Review of Condensed Matter Physics, {\bf 1}, 27 (2010).

\bibitem{Paglione2010}%review
Johnpierre Paglione and Richard L. Greene, Nature Physics {\bf 6}, 645 (2010).

\bibitem{Margadonna2009}%FeSe pressure
S. Margadonna, Y. Takabayashi, Y. Ohishi, Y. Mizuguchi, Y. Takano, T. Kagayama, T. Nakagawa, M. Takata, and K. Prassides, Phys. Rev. B {\bf 80}, 064506 (2009).

\bibitem{Guo2010}%KFe2Se2
J. Guo, S. Jin, G. Wang, S. Wang, K. Zhu, T. Zhou, M. He, and X. Chen, Phys. Rev. B {\bf 82}, 180520 (2010).

\bibitem{Ying2010}%CsFe2Se2
J. J. Ying, X. F. Wang, X. G. Luo, A. F. Wang, M. Zhang, Y. J. Yan, Z. J. Xiang, R. H. Liu, P. Cheng, G. J. Ye, and X. H. Chen, arXiv:1012.5552 (2010).

\bibitem{Li2010}%RbFe2Se2
C. H. Li, B. Shen, F. Han, X. Zhu, and H. H. Wen, arXiv:1012.5637 (2010).

\bibitem{Fang2010}%TlFe2Se2
M. Fang, H. Wang, C. Dong, Z. Li, C. Feng, J. Chen, and H. Q. Yuan, arXiv:1012.5236 (2010).

\bibitem{Shermadini2011} Z. Shermadini, A. Krzton-Maziopa, M. Bendele, R. Khasanov, H. Luetkens, K. Conder, E. Pomjakushina, S. Weyeneth, V. Pomjakushin, O. Bossen, A. Amato, arXiv:1101.1873 (2011).

\bibitem{Pomjakushin2011}%KFe2Se2 neutron
V. Yu. Pomjakushin, E. V. Pomjakushina, A. Krzton-Maziopa, K. Conder, and Z. Shermadini, arXiv:1102.3380 (2011).

\bibitem{Bao2011}%KFe2Se2 neutron and high T study
W. Bao, G. N. Li, Q. Huang, G. F. Chen, J. B. He, M. A. Green, Y. Qiu, D. M. Wang, J. L. Luo, and M. M. Wu, arXiv:1102.3674 (2011).

\bibitem{Ryan2011}%Mossbauer
D. H. Ryan, W. N. Rowan-Weetaluktuk, J. M. Cadogan, R. Hu, W. E. Straszheim, S.L. Bud'ko, and P.C. Canfield, arXiv:1103.0059 (2011). 

\bibitem{Hu2011}%KFe2Se2 Ames sample growth and characterization
R. Hu, K. Cho, H. Kim, H. Hodovanets, W. E. Straszheim, M. A. Tanatar, R. Prozorov, S. L. Bud'ko, P. C. Canfield, arXiv:1102.1931 (2011).

\bibitem{Altarawneh2008}%BaKFe2As2 Hc2 TDO
M. M. Altarawneh, K. Collar, C. H. Mielke, N. Ni, S. L. Bud'ko, and P. C. Canfield, Phy. Rev. B {\bf 78}, 220505(R) (2008).

\bibitem{Ni2008}%BacoFe2As2 Hc2 Resistance
N. Ni, M. E. Tillman, J.-Q. Yan, A. Kracher, S. T. Hannahs, S. L. Bud'ko, and P. C. Canfield, Phys. Rev. B {\bf 78} 214515 (2008).

\bibitem{Yuan2009}%BaKFe2As2 Hc2 Resistance
H. Q. Yuan, J. Singleton, F. F. Balakirev, S. A. Baily, G. F. Chen, J. L. Luo and N. L. Wang, Nature(London) {\bf 457}, 565 (2009).

\bibitem{Kano2009}%BaCoFe2As2 Resistance
M. Kano, Y. Kohama, D. Graf, F. Balakirev, A. S. Sefat, M. A. McGuire, B. C. Sales, D. Mandrus, and S. W. Tozer, J. Phys. Soc. Jpn. {\bf 78} 084719 (2009).

\bibitem{Mielke2001}%PDO
C. Mielke, J. Singleton, M.-S. Nam, N. Harrison, C. C. Agosta, B. Fravel, and L. K. Montgomery, J. Phys.: Condens. Matter {\bf 13}, 8325 (2001).

\bibitem{Coffey2000}%PDO
T. Coffey, Z. Bayindir, J. F. DeCarolis, M. Bennett, G. Esper, and C. C. Agosta, Rev. Sci. Instrum. {\bf 71}, 4600 (2000).

\bibitem{Altarawneh2009}%PDO
M. M. Altarawneh, C. H. Mielke, and J. S. Brooks, Rev. Sci. Instrum. {\bf 80}, 066104 (2009).

\bibitem{Mizuguchi2011}%KFe2Se2 Hc2 DC
Y. Mizuguchi, H. Takeya, Y. Kawasaki, T. Ozaki, S. Tsuda, T. Yamaguchi, and Y. Takano, Appl. Phys. Lett {\bf 98} 042511 (2011).

\bibitem{Ando1999}%Hc2 definition cuprates
Y. Ando, G. S. Boebinger, A. Passner, L. F. Schneemeyer, T. Kimura, M. Okuya, S. Watauchi, J. Shimoyama, K. Kishio, K. Tamasaku, N. Ichikawa, and S. Uchida, Phys. Rev. B {\bf 60}
12475 (1999).

\bibitem{Wang2011} %KFe2Se2 Hc2 DC
D. M. Wang, J. B. He, T.-L. Xia, G. F. Chen, arXiv:1101.0789 (2011).

\bibitem{Werthamer1966}%Hc2(0) WHH theory
N. R. Werthamer, E. Helfand, and P. C. Hohemberg, Phys. Rev. {\bf 147}, 295 (1966).

\bibitem{Clogston1962}%Pauli limit
A. M. Clogston, Phys. Rev. Lett. {\bf 9}, 266 (1962).

\bibitem{Chandrasekhar1962}%Pauli limit
B. S. Chandrasekhar, Appl. Phys. Lett. {\bf 1}, 7 (1962).

\bibitem{Maki1966}%Pauli limit
K. Maki, Phys. Rev. {\bf 148}, 362 (1966).

\bibitem{Poole2000}%SC Handbook
Charles P. Poole, Jr. (editor) {\it Handbook of Superconductivity} (Academic Press, San Diego) (2000).

\bibitem{Khim2010}%Hc2 FeTe0.6Se0.4
S. Khim, J. W. Kim, E. S. Choi, Y. Bang, M. Nohara, H. Takagi, and K. H. Kim, Phys. Rev. B {\bf 81}, 184511 (2010).

\bibitem{Fang2010R}%Hc2 Fe1.11Te0.6Se0.4
M. H. Fang, J. H. Yang, F. F. Balakirev, Y. Kohama, J Singleton, B. Qian, Z. Q. Mao, H. D. Wang and H. Q. Yuan, Phys. Rev. B {\bf 81}, 020509(R) (2010).

\end{thebibliography}
\end{document}